%% file: ms7.tex
\documentclass[12pt,preprint]{aastex}

\newcommand{\bdv}[1]{\mbox{\boldmath$#1$}}

\def\au{{\rm AU}}

\def\mas{{\rm mas}}

\def\max{{\rm max}}
\def\min{{\rm min}}
\def\rel{{\rm rel}}

\def\e{{\rm E}}
\def\bpi{{\bdv\pi}}
\def\bmu{{\bdv\mu}}

\begin{document}
\title{The Korea Microlensing Telescope Network (KMTNet) Alert 
Algorithm and Alert System}

\input author.tex

\begin{abstract}
We describe a new microlensing-event alert algorithm that is tailored
to the Korea Microlensing Telescope Network (KMTNet) multi-observatory
system.
The algorithm focuses on detecting ``rising'' events, i.e., events whose brightness is increasing as a function of time.
  The algorithm proceeds in three steps. It first identifies 
light curves with at least $N_{\rm high}$ points that are at least $3\sigma$ 
above the median and that had been taken since shortly before
the previous search for new events.  It
then demands that there are at least $N_{\rm high}$ {\it consecutive} 
high points when considering
any combination of light curves from one, two, or three observatories.  
Finally, it
fits to a ``rising microlensing model'' consisting of a broken line,
i.e., flat before some time $t_{\rm rise}$ and rising linearly afterward.
If this fit is better than a flat line by 
$\Delta\chi^2>\Delta\chi^2_{\rm thresh}$, the
light curve is sent for human review. Here, 
$(N_{\rm high},\Delta\chi^2_{\rm thresh})=(5,250)$ or (10,400), 
depending on the field cadence.
For 2018, KMTNet alerts will initially be restricted to a few northern
bulge fields and may gradually extend to the full northern bulge.
Further expansion of coverage is expected in 2019.

\end{abstract}


\section{{Introduction}
\label{sec:intro}}

Real-time microlensing alerts have played a crucial role in the
development of the field.  Prior to the discovery of the first
microlensing event, \citet{gouldloeb} had already proposed real-time
alerts as a means to bridge the gap between the technical requirements
of finding microlensing events and those of monitoring these events with
sufficient cadence to identify and characterize planetary signatures.
Even toward the densest star fields of the
Galactic bulge, the microlensing event rate is only 
$\gamma\sim 10^{-5}\,{\rm yr}^{-1}$.  Hence, one must monitor tens 
of square degrees, with typical stellar surface densities of
$\sim 5\times 10^6\,{\rm deg}^{-2}$ (to limiting depth $I\la 20$)
to find of order $10^3$ microlensing events per year.  With the camera
fields of view available (or foreseen) at that time, this required tiling 
the target area once or at most a few times per night.  These low
cadences $\Gamma\sim 1\,{\rm day}^{-1}$ were quite adequate to find 
microlensing events, with their characteristic timescales $t_\e\sim 20\,$day, 
because they still yielded $N=\Gamma t_\e\sim 20$ epochs per Einstein
crossing time.  However, this cadence is inadequate to characterize
all but the largest planets because the timescales of the planetary
deviations are typically shorter than the events by 
$t_{\rm pert}\sim \sqrt{q}t_\e$, where $q$ is the planet/host mass ratio.
Hence, to obtain, say, five epochs per perturbation timescale
(i.e., 10 over the whole perturbation) requires a higher cadence by a 
factor $q^{-1/2}/4$.  This corresponds 
$\Gamma\sim (0.4,1,4)\,{\rm hr}^{-1}$ for (Saturn, Neptune, Earth) mass planets
(assuming that the typical $t_\e\sim 20\,$day timescale corresponds to
an $M\sim 0.5\,M_\odot$ host).  Hence, \citet{gouldloeb} advocated
intensive follow-up observations from multiple sites of individual
microlensing events, which of course required that these events
be identified and publicly announced while they were still ongoing
and therefore sensitive to planets.

Such real-time alert systems were quickly established by the 
MACHO \citep{machoalert} and OGLE \citep{ews1} collaborations.  These alerts,
and their further development by OGLE \citep{ews2,ogle-iv}, 
became the basis for follow-up teams such as PLANET \citep{planet}, GMAN 
\citep{gman}, and MPS \citep{mps},
to conduct the first microlensing planet searches and so to find one of the
first microlensing planets, OGLE-2005-BLG-390Lb \citep{ob05390}.
Later, MOA began conducting a
new survey and likewise almost immediately started to issue
its own independent alerts \citep{moaalert}.  New followup groups formed, with 
$\mu$FUN \citep{gould10} in particular organizing itself around the new 
strategy proposed by \citet{griest98} 
of primarily monitoring high-magnification 
events because these had the highest sensitivity.  This channel likewise
led to one of the first microlensing planet discoveries, OGLE-2005-BLG-071
\citep{ob05071}.  As we will discuss,
this development was of particular note in the present context because
events reaching high-magnification $A_\max\gg 1$ only remain near peak for
$\sim 3t_\e/A_{\rm max}$ during which time they should be monitored at
very high cadence to fully exploit their high sensitivity to planets.

Beginning in 2014, a completely new application of microlensing alerts
emerged: {\it Spitzer} microlensing 
\citep{prop2013,prop2014,prop2015a,prop2015b,prop2016,prop2017}.  
By combining ground-based
light curves with those obtained by {\it Spitzer} in solar orbit
at $\sim 1\,\au$, one can measure the so-called microlens parallax $\bpi_\e$,
\begin{equation}
\bpi_\e \equiv {\pi_\rel\over\theta_\e}{\bmu_\rel\over\mu_\rel}
\qquad
\theta_\e^2\equiv \kappa M\pi_\rel
\qquad
\kappa \equiv {4 G\over c^2\,\au}\simeq 8.14\,{\mas\over M_\odot},
\label{eqn:piedef}
\end{equation}
where $(\pi_\rel,\bmu_\rel)$ are the lens-source relative 
(parallax, proper motion), $\theta_\e$ is the angular Einstein radius,
and $M$ is the lens mass \citep{refsdal66,gould94,gouldhorne}.
This measurement, by itself, provides
a powerful constraint on the physical parameters $M$ and $\pi_\rel$
\citep{han95}, and, whenever $\theta_\e$ is also measured, directly
determines these quantities:
\begin{equation}
M = {\theta_\e\over\kappa\pi_\e};
\qquad
\pi_\rel = \theta_\e\pi_\e .
\label{eqn:mpirel}
\end{equation}
Because {\it Spitzer} is a narrow-angle instrument, it must be directed
to point at {\it ongoing} microlensing events.  This requires microlensing
alerts.  Moreover, in contrast to high-magnification microlensing
alerts, which can be profitably exploited with even a few hours notice
before the event peaks, {\it Spitzer} alerts require a long lead time
simply because spacecraft commands are uploaded only once per week and
must be delivered to {\it Spitzer} operations three days before upload.
See \citet{yee15} and Figure~1 of \citet{ob140124}.

Because of this long lead time, the overwhelming majority $(>90\%)$
of {\it Spitzer} alerts have come from OGLE, whose excellent photometry
and well-oiled alert machinery enable highly reliable microlensing alerts
long before peak when the events are still usually very faint.

The Korea Microlensing Telescope Network (KMTNet, \citealt{kmtnet})
is presently the largest microlensing survey, but up until this
point has not issued microlensing alerts.  KMTNet consists of three
identical 1.6m telescopes at CTIO (Chile, KMTC), SAAO (South Africa, KMTS) 
and SSO (Australia, KMTA), each equipped with a $4\,{\rm deg}^2$ camera.
This enables 24 hour coverage (weather permitting) of sky areas
$\Omega= (12,41,85,97)\,{\rm deg^2}$ at cadences 
$\Gamma\geq (4,1,0.4,0.2)\,{\rm hr}^{-1}$, which are adequate to detect
and characterize roughly (Earth, Neptune, Saturn, Jupiter) mass planets
without followup observations.  See Section 2 of \citet{2016ef} for more
details on cadence as a function of area.

For this reason, microlensing alerts were initially a very low
priority of KMTNet.  Instead, after more basic tasks of commissioning
the telescope and data pipelines were completed, the most urgent task
was to develop a completed-event finder, which could identify
microlensing events {\it after the season was completed}
\citep{eventfinder,2016k2,2016ef}.  A microlensing event always leads
to a temporary increase in brightness of the source. As such, it is
characterized by at least three features: an increase in brightness
relative to a baseline flux level, a
peak, and a decrease in brightness relative toward the baseline.
The event-finder algorithm is focused on detecting ``completed
events,'' which have undergone all three of those phases.

For 2016 data, for example, these post-season events have yielded planets
for three KMTNet-only events \citep{kb160212,kb161107,kb161397}, while
also contributing critical data on planetary signals in
events discovered by others 
\citep{ob160596,ob161195,ob161190,ob160263,ob160613,ob161067}.

In contrast, an alert-finder algorithm would be required to detect
events in progress without the benefit of all features. For example, a
"rising event" is one that is before peak and for which the flux is
simply increasing as a function of time. Thus, the algorithmic
constraints are different than for the event-finder.

As discussed in some detail by \citet{eventfinder}, there are
several reasons for KMTNet to develop an alert system.  First, in
contrast to its original design strategy of observing just $16\,{\rm deg}^2$
at $\Gamma=6\,{\rm hr}^{-1}$, which would have densely covered all caustic
crossings whenever KMTNet was observing, the great majority of the area
covered in the new strategy has cadences $\Gamma\leq 1\,{\rm hr}^{-1}$.
This means that many caustic crossings, which typically last 1--2 hours,
will be missed unless followup observations are organized for their
predicted times.

Second, as mentioned above, the relatively rare high-magnification
events can always benefit from higher cadence.  This is particularly
true for the new strategy because at $\Gamma\leq 1\,{\rm hr}^{-1}$, it
is rather unlikely that planetary perturbations at high magnification
can be properly characterized.  However, even under the old KMTNet observing 
strategy, there was a substantial chance that high-magnification events would
not be fully characterized, both because $\Gamma=6\,{\rm hr}^{-1}$
is still relatively low for these events and because high-magnification
events are more likely to be saturated $(I\la 13$) or at least in the
non-linear regime $(I\la 13.5)$ in KMTNet data.

Finally, KMTNet alerts could potentially contribute to finding {\it Spitzer}
targets.  This is particularly true in the northern Galactic bulge
where extinction is generally higher and where OGLE generally observes
at much lower cadences than KMTNet.  Higher extinction means that
at fixed optical brightness, the sources are brighter in {\it Spitzer},
making them more favorable targets.  At least as judged by 
the ``completed-event finder'' results for the BLG14 field for the 
2016 season\footnote{http://kmtnet.kasi.re.kr/ulens/event/2016/}, 
KMTNet finds many more events in the
northern bulge compared to OGLE, which is not the case in the southern
bulge.

\section{{Scientific Objectives and Practical Constraints}
\label{sec:obj+const}}

Ideally, one would like to announce microlensing alerts for
all events that can be distinguished from ``noise''
(whether instrumental or astrophysical) based on a survey's data, and
to do so as soon as the data in which they are first distinguishable from noise
are taken.  This goal is in fact very nearly achieved by the MOA
alert system: images are reduced on site within minutes after they
are taken, and new sources are automatically identified and shown to 
team members.  Then, after team review, which can be quite rapid
in urgent cases, the events are publicly announced.  

However,
a variety of practical constraints prevent such an approach from 
being applied to KMTNet data, most of them related to the fact that
the KMTNet system has roughly 12 times more pixels than MOA. 
Due to the sheer volume of data from KMTNet, the practical
constraints due to finite resources must be taken into account in
both the design and implementation of an alert-finder algorithm. 
Therefore,
in devising an alert system, it is essential to begin by prioritizing
scientific objectives and weighing these against practical constraints.

\subsection{{Scientific Objectives}
\label{sec:obj}}

The only science objective for the KMTNet alert system is to enable
followup observations of ongoing microlensing events, either by {\it Spitzer}
or from the ground.  The OGLE and MOA alerts, by contrast, are also
the principal means by which these groups publicly identify microlensing events.
However, for KMTNet, this latter function is served by the event-finder, which
finds events that are
too faint to be detected by any reasonable alert algorithm.  

Nevertheless, although the science objective appears unitary, it still
must be further parsed in order to be properly weighed against
practical constraints.  As outlined in Section~\ref{sec:intro}, follow-up
observations can be triggered in pursuit of several different goals,
with the main ones being
\begin{enumerate}
\item[]{1) Monitoring by {\it Spitzer}}
\item[]{2) Monitoring high-magnification events}
\item[]{3) Monitoring identified planets}
\item[]{4) Monitoring predicted caustic crossings}
\end{enumerate}

The first two applications virtually require that the alert be issued
before peak.  In the case of high-magnification events, this is an 
absolute requirement.  For {\it Spitzer} observations the requirement
is ``almost absolute'' because the minimum 3-day delay between upload
and first observation, combined with the fact most events peak earlier
as seen from {\it Spitzer} than from Earth, together imply dramatically
reduced probability of measuring the microlens parallax for events
that are alerted after peak.

The last two applications require only that the alert be issued prior
to the appearance (or, strictly speaking, the completion) of the
anomaly {\it and} that this anomaly be recognizable in the publicly 
available light curves that can be accessed from the alert webpage.
These applications clearly favor alerts before peak (as half of
all anomalies occur before peak), but they do not absolutely require it.

\subsection{{Practical Constraints}
\label{sec:const}}

The main practical constraints all basically derive from the volume
of data combined with the limited computational and human resources
to deal with these data.  KMTNet makes approximately $10^{12}$ photometric
measurements per season, distributed over $5\times 10^8$ catalog stars
within a total of 27 fields of $4\,{\rm deg}^2$.  Each of its three
cameras has $3.4\times 10^8$ pixels distributed over four chips.
Note that slightly less than 10\% of all KMTNet observations are
in $V$ band.  These are not used for alerts (nor for post-season
event-finding) and so are not reduced in real time.

The first constraint is that the photometric pipeline must be able to
``keep up'' with the data flow.  That is, the data should be
processed at a rate that is equal to or faster than the mean rate at
which they are acquired.  We have placed ``keep up'' in quotation
marks because there will inevitably be some delay between the time
that the data are taken and the time that they are reduced and archived, 
which we discuss in some detail further below.  However, what is important
in the present context is that this delay be small compared to the
duration of a typical microlensing event.  That is, the reductions
must be basically concurrent with the observations and, in particular,
at the height of the microlensing season this requirement would
be that the delay cannot increase
due to increasing backlog of unprocessed observations.  While this
constraint may perhaps seem too obvious to even mention, in fact,
the KMTNet pipeline only began catching up with the observations in mid-May
2018.  For example, the reductions of 2017 KMTNet data did not
begin until December 2017 and were not completed until three months later.
See the discussions in \citet{eventfinder,2016k2}.
Moreover, the onset of reductions of 2018 data were delayed by 2 months
due to a massive upgrade of the star catalog \citep{2016ef}, and by
the evaluation of the statistical characteristics of these 
catalog stars, which is required for efficient real-time selection of
microlensing events (Section~\ref{sec:tworeq}).  Hence, prior to 2018, 
an alert system would not have been possible.  Moreover, as we will
describe in Section~\ref{sec:ramp}, 
the problems engendered by this ``catching up'' will both
impact the inaugural phase of KMTNet alerts and are likely to impact the
continued functioning of this system.

The second constraint is
a problem of data input/output (I/O) rates and storage. In order to 
reliably detect microlensing events, all $10^{12}$ measurements (or
whatever fraction have been acquired by a particular date) must be
readily available on storage disks local to the computers
applying the algorithm. This requires large hard disks. In
principle, the required storage could be reduced using a tiered system, 
in which a subset (say the previous month)
of the data has immediate local availability and the rest can be acquired
from more distant storage at a more leisurely pace for the handful of events 
that are identified
from the recent-month of data as ``interesting''.  In practice, such
a tiered system would be unwieldy and prone to failures.  As we will
show below, 8 bytes are required to store one measurement.  Hence,
simply to access the data requires 8 Terabyte of rapid-access memory.
In fact, KMTNet possesses eight 2-Terabyte solid state disks (SSD), two
attached to each server that is responsible for processing one of the
four chips on the camera.  Hence, this requirement is 
satisfied by the KMTNet system.  In particular, no tiered system is
required.

The third constraint is that at some point, a human must review all
candidate alerts to assess the probability that they are real
microlensing events. This affects both the relevant timescales (below)
and the efficiency and reliability requirements of the algorithm.

The remaining constraints all relate to the timescale for the
complete alert-finder process to run from start to finish, i.e.,
the turnaround time from the
acquisition of photometric measurements to the human review of updated 
light curves that determines which light-curve variations should be announced
as ``clear'' or ``probable'' microlensing.  Whichever step takes the longest
will determine this turnaround time, which in turn will enable and/or limit
the scientific applications.

For KMTC and KMTA, images are transferred from the camera to Korea within
a few minutes of the time they are taken.  Due to lower band width from
South Africa, the arrival of the final KMTS image for a given night
can be delayed by up to 6 hours
after it is taken, depending on a variety of factors but mainly the season.
In particular, $\sim 6$ hour delays are typical for the entire peak season
from May through August.

Once they arrive, images are pre-processed (cross-talk correction, de-biasing, 
flat-fielding, bad-pixel masking, and writing to disk) in about two minutes
on average, using eight processors that are reserved for this purpose.
Aligning and subtracting images requires 4 minutes, and DIA photometry
requires 5 minutes.  Another 1 minute is required to append these
photometric measurements to the individual light-curve files. 
Therefore, for KMTC and KMTA, the total time from image acquisition to
photometry is well under one hour, but for KMTS these processes often 
require up to seven hours due to the delay in transferring the data.

These constraints imply that there are three potential timescales that
are relevant for the machine review (i.e., machine-based vetting
prior to human review) of light curves: one hour, 12 hours, 
and 24 hours.  That is, machine reviews taking substantially less than one 
hour
are not useful because the basic reductions already take a significant
fraction of an hour.
Machine reviews taking 12 hours would enable two human reviews per day,
one at about UT 21:00, i.e., roughly 12 hours after the arrival of
KMTS and KMTC data, and one at UT 07:00, i.e., roughly 12 hours after the
arrival of KMTA data.  Conveniently, these would be at the beginning
and end of the Korean work day.  Finally, an alert-finder algorithm 
taking less than 24 hours would permit daily human reviews, which would
be well matched to the human diurnal cycle.

\section{{Alert Algorithm}
\label{sec:algorithm}}

From a conceptual standpoint, the alert-finder algorithm works as follows:

Every light curve is examined by machine to determine whether it is
consistent with a rising microlensing event.  It is checked to see if
there is an increase in brightness over the baseline level, i.e., by
evaluating whether there have been at least $N_{\rm high}$ recent consecutive
points that were more than 3-sigma above the median.  Then, those light
curves with such a sustained increase in brightness are fitted with a ``rising
microlensing light-curve model'' and compared to a flat line, i.e., by
evaluating $\Delta\chi^2$ relative to a $\Delta\chi_{\rm thresh}^2$ threshold. 
Events passing those metrics are vetted for duplicates and
variables. Finally, they are reviewed by a human operator to
determine which events are real microlensing events. The entire
process prior to review by a human operator constitutes a ``machine
review.''

However, as discussed in Section \ref{sec:const}, there are a number
of practical considerations that affect the implementation of this
concept. The biggest one is the enormous data volume compared to
limited computational and human resources. There are also complexities
introduced by the fact that the data come from three distinct sites,
each of which have different observing properties and different rates
of data transfer. We discuss the practical implementation of the
algorithm and its implications below.

\subsection{{Qualitative Characteristics}
\label{sec:quality}}

The considerations described in Sections~\ref{sec:obj} and \ref{sec:const}
imply that the machine review should focus on and robustly identify
rising (i.e., pre-peak) events, should operate
as rapidly as possible, and should minimize the number of false positives
in order to permit rapid human review.  Each of these ``positive'' goals
implies some ``negative'' cost.

``Focusing on rising events'' implies possibly foregoing the detection of
events that have already peaked.  Because all events must rise before
they peak, it may appear that this is no cost at all\footnote{Here, we
are ignoring events that peak between seasons.  We address these
in Section~\ref{sec:ramp}.}.  However, low-amplitude
events may not contain enough information for reliable detection before
peak and yet may acquire such information after peak.  The choice being
made is to forgo these events as alerts.  First, these events will be
found at the end of the season by the event-finder.  Second, as discussed
in Section~\ref{sec:obj}, most of the scientific driver for alerts comes
from pre-peak alerts.  Third, events that are so noisy that they cannot
be reliably identified before peak are unlikely to give rise to
recognizable anomalies that lead to followup observations after peak.

``Operating as rapidly as possible'' means minimizing the number
of data points examined for
each light curve and then, on this basis, selecting only a small
fraction of all light curves for more intensive analysis.  This implies,
in particular, that noisy light curves whose microlensing signature
is only recognizable by averaging over a very large number of points
will be missed.  Again, the rationale for this choice is that such
events will be found by the event-finder at the end of the season and
that such events are much less likely to give rise to identifiable anomalies
that would trigger followup observations.

``Minimizing the number of false positives'' implies setting the
machine threshold high enough that some real events will be missed
and hence will not be shown to the operator.  By contrast, the event-finder
has a relatively low machine (event-detection)
threshold.  However, in that case, human
review is required only once per year, not once or several times per day.

\subsection{{Two Machine-Review Requirements to Trigger Human 
Review}
\label{sec:tworeq}}

Based on the qualitative characteristics described in 
Section~\ref{sec:quality}, we demand that a light curve
satisfy the following two quantitative requirements to pass on the
stage of human review:
\begin{enumerate}
\item[]{A1) At least $N_{\rm high}$ ``consecutive'' flux  measurements that are
 $\geq 3\,\sigma$ above the median}
\item[]{A2) $\Delta\chi^2>\Delta\chi^2_{\rm thresh}$ improvement for 
a ``rising microlensing light-curve model''}
\end{enumerate}
Note that requirement (A1) preferentially suppresses (though it does not
eliminate) variable stars because their ``scatter'' is typically dominated
by astrophysical (rather than Poisson or instrumental) variations.
The practical implementation of these two conditions will be described in
Section~\ref{sec:implement}, but first, several terms within these conditions 
require further explanation.

The term ``consecutive'' is placed in quotation marks because its definition
is fairly complex.  First, we mask any point that fails to meet any one
of the following three conditions
\begin{enumerate}
\item[]{B1) Background light below the 92nd percentile}
\item[]{B2) Seeing below the 84th percentile}
\item[]{B3) $\chi^2$(DIA-PSF fit) below 100}
\end{enumerate}
These three statistical quantities, as well as the median and $\sigma$
of the flux from (A1), are all derived from previous seasons 
(currently the 2016 season is the reference) and are
observatory-specific and specific to this particular star.  Note also that
the ``$\sigma$'' in (A1) is actually one-half of the difference between
the 84th percentile and the 16th percentile of the previous
  seasons' flux (i.e., it is not the photometric uncertainty).
These statistics are
cataloged and are not recomputed for each application of the algorithm.

Second, the $N_{\rm high}$ points can be ``consecutive'' within any
combination of the $3N_{\rm fields}$ light-curve files.  That is, the
event may be observed in several (``$N_{\rm fields}$'')
overlapping fields, and for each
of these it is observed from three observatories.  There are
$i=1, \ldots, (2^{3N_{\rm fields}}-1)$ combinations of  light-curve files
(so $N_{\rm fields}=1$ has 7 combinations including, for example,
  KMTC alone or in combination with KMTS and/or KMTA.).
For each of these, we write
$i$ as a binary number with $3N_{\rm fields}$ digits (i.e., each
  digit represents a light curve, which is either under consideration
  or not).
Then, in asking whether there are $N_{\rm high}$ consecutive points,
we consider only the light-curve files for which the entry in this
binary representation is ``1'', and ignore files for which the entry
is ``0''.
For example, this number is ``101'' for $i=5$ and $N_{\rm fields}=1$,
  and only two out of three light-curve files are considered.

The ``rising microlensing light-curve models'' for the flux $F$ are defined as
\begin{equation}
F(t) = a_0 + a_1 (t-t_{\rm rise})\Theta(t-t_{\rm rise}),
\label{eqn:rise}
\end{equation}
where $\Theta$ is a Heaviside step function and $a_1$ is constrained to
be positive.  Note that 
Equation~(\ref{eqn:rise}) has only one non-linear parameter ($t_{\rm rise}$).
Hence, a good estimate of the best fit can be obtained by carrying out a 
two-parameter $(a_0,a_1)$ linear fit on a one-dimensional discrete
set of models characterized by different $t_{\rm rise}$.  The $\Delta\chi^2$
improvement is then measured relative to a flat model, i.e., $F(t) = a_0$.

\subsection{{Practical Implementation}
\label{sec:implement}}

Before assessing how well these two machine-review requirements (A1) and (A2) 
satisfy
the scientific objectives and practical constraints, it is first
necessary to describe their practical implementation.

Because of the data volume, the system for storing light-curve data
substantially affects the speed of the alert-finder algorithm. For
example, it is significantly faster to access information stored in
binary rather than in ASCII format. Thus, in discussing practical
implementation, we must begin by describing how light-curve data is
stored.

Light curves are organized by ``patches''.  There are 64 slightly overlapping
patches per chip, each of which is $1\,{\rm deg}^2$.  Within each patch,
all the light curves have exactly the same length and exactly the
same epochs.  If a particular photometric measurement fails for any
reason, it is flagged but left in the file.  Therefore, there is only
one date file (and also only one background-light file) for each patch,
which are read into memory at the start of processing the patch.  The remaining
variables describing each photometric measurement are the
difference flux, the flux error, the $\chi^2$ of the DIA fit, and the
seeing.  These require a total of $4\times 2=8$ bytes in the direct-access
binary-format files.  The flux $F$ is stored as the
nearest (2-byte) integer of a modified ``Lupton magnitude'',
$K_F = 3000\sinh^{-1}(F/3000)$, the flux error is stored as the 
nearest (2-byte) integer, and the remaining two parameters are multiplied
by 100 and then stored as 2-byte integers.  The flux units are in ADU,
with 1 ADU corresponding to $I\simeq 28$.  Therefore, storing the
modified Lupton magnitude and the flux error to the nearest integer
does not induce any significant loss of precision.

The vetting program has three input parameters, the $\chi^2$ threshold
$\Delta\chi^2_{\rm thresh}$, the number of consecutive
``high'' points ($N_{\rm high}$) required to trigger a deeper investigation,
and the last date of data reviewed in the previous search (which, to first
approximation,
is set as the earliest of the previous upload dates from the three
observatories, but see below).  
We adopt $\Delta\chi^2_{\rm thresh}=400$ for fields 
with $\Gamma\geq 1\,{\rm hr}^{-1}$ cadence and $\Delta\chi^2_{\rm thresh}=250$
for fields with $\Gamma\leq 0.4\,{\rm hr}^{-1}$ cadence.  For the first,
we find from
practical tests that very few candidates can be reliably identified
as even ``probable microlensing'' (i.e., the minimum requirement
for them to be publicly announced) at $\Delta\chi^2<400$, and of those
that can be, almost none are identifiable at $\Delta\chi^2<350$.
Moreover, extremely few of those that could be identified at $\Delta\chi^2<400$
and do not reach $\Delta\chi^2>400$ within a day or so, prove to be worthy
of followup observations.  However, for the lower-cadence fields, we find
that reliable alerts (i.e., ``probable'' or ``clear'' microlensing) are
often possible at or just above $\Delta\chi^2=250$.

We choose $N_{\rm high}=10$ for fields with $\Gamma\geq 1\,{\rm hr}^{-1}$ cadence
and $N_{\rm high}=5$ for fields with $\Gamma\leq 0.4\,{\rm hr}^{-1}$ cadence.
Below these thresholds, we find that the number of spurious candidates
rises dramatically.  See Figure~12 of \citet{eventfinder} for the cadences 
of each field.

The last parameter (last date of review)
is, of course, updated for each operation of the program.  
In fact, the ``last date of review'' is a rather complex quantity, in part
because it is actually the date of the last observation available
at the time of the review. Thus,
the three different observatories necessarily have different 
``last dates''.  Nor is it a viable solution to simply keep track of these
three dates separately.  In order to consider strings of consecutive
points from multiple observatories, one must have a common ``last date''.
We note that there is no fundamental harm in setting this date arbitrarily
far back in the past: this only increases operation time.  Therefore
we set this date conservatively to make sure that no instances of $N_{\rm high}$
consecutive points are missed.  We first evaluate the last date for
each observatory, $t_{j,\rm last},j=1,2,3$.  We then set
\begin{equation}
t_{\rm last} = \max[(t_{\rm now} - 4\,{\rm days}),\min(t_{j,\rm last})]
\label{eqn:tlast}
\end{equation}
where $t_{\rm now}$ is the most recent observation from any observatory.
The first condition ($t_{\rm last} \geq (t_{\rm now} - 4\,{\rm days})$) guards
against duplicating the same searches every day for cases of long spans
of bad weather at one observatory.  Finally, we set
$t_{\rm last} = (t_{\rm now} - 4\,{\rm days})$ for $t_{\rm now}$ within one
day of the full moon in order to guard against gaps in usable data
due to many consecutive data points that violate condition (B1).

At the present time, we analyze the three sites together but 
each field separately, i.e., $N_{\rm fields}=1$,
for reasons that we discuss in Section~\ref{sec:ramp}.

The two requirements (A1) and (A2) are implemented through three steps, with 
the first two steps corresponding to the first requirement and the
last step corresponding to the second.

\subsubsection{{Step 1: Vetting for $N_{\rm high}$ points}
\label{sec:ngoodvet}}

The program begins by stepping through each data file, starting at the
``bottom'' (last data entry) and including $N_{\rm high}+10$ points
before $t_{\rm last}$.  It 
counts the number of points from all data files combined
that satisfy the three conditions B1--B3, and for which the flux
is at least $3\sigma$ above the median. If this number is at least
$N_{\rm high}$ then it moves on to the second step.

The great majority of light curves fail this test.  Because the test
is simple and typically requires reading only a few dozen points,
it is very fast.

\subsubsection{{Step 2: Vetting for $N_{\rm high}$ consecutive points}
\label{sec:ngoodconsec}}

A light curve that passes Step 1, is then fully tested for requirement (A1).
First, an internal list is created that consists
only of the points identified in Section~\ref{sec:ngoodvet} 
that satisfy conditions B1--B3.
The list combines data from all three files
and is ordered by date, and with the data file from which the data point
was drawn marked.  Data points that are at least $3\sigma$ above the
median are flagged as ``high'',
while the others are ``not high''.  The program then goes through
this list $i=1, \ldots, 2^{N_{\rm files}}-1$ times as described in 
Section~\ref{sec:tworeq} to check if any combination of files has
$N_{\rm high}$ consecutive ``high'' points.

A substantial majority of stars that pass Step 1, fail Step 2.

\subsubsection{{Step 3: Vetting for rising microlensing light curve}
\label{sec:rising}}

Stars that pass Step 2 are checked to see if they have ``rising microlensing
light curves'' as characterized by Equation~(\ref{eqn:rise}).  For
this purpose, the {\it entire} light curve for the season is read into
memory, i.e., typically of order 100 times more data than in Steps 1 and 2.
Again, data points failing to satisfy conditions B1--B3 are flagged
and ignored.  Then a series of $k=1,\dots N_{\rm fit}=16$ fits are carried
out, which are defined by
\begin{equation}
t_{\rm rise} = t_{\rm now} - 2^{(k-3)/2}\,{\rm day},
\label{eqn:trise}
\end{equation}
where $t_{\rm now}$ is the most recent observation.  Hence,
$(t_{\rm now}-t_{\rm rise})$ spans a range from 0.5 to 90.5 days.
The fitting procedure is similar to the one described 
by \citet{eventfinder} for their much 
more extensive set of models ($\sim 5000$ versus 16).  Each fit
is carried out four times.  First, all the data (from all
$N_{\rm files}=3N_{\rm fields}$ files) are fitted to derive the $a_{m,j}$
($m=0,1$, $j=1,\ldots N_{\rm files}$), i.e., 2$N_{\rm files}$ parameters.
Second, the $\chi^2$ contribution of each data point is evaluated.
The 5\% worst data points from each file are eliminated.  Third,
all data files are refitted with these points removed in order to
redetermine the $a_{0,j}$ and $a_{1,j}$.  
Fourth, the $\chi^2_{j,\rm 2-parm}$ of each data file is evaluated.

The same surviving data points (from the second fit) are then fitted
to a flat line, $F_j(t) = a_{0,j}$, $\chi^2_{j,\rm 1-parm}$ is evaluated for each
data file, and hence $\Delta\chi^2_j=\chi^2_{j,\rm 1-parm}- \chi^2_{j,\rm 2-parm}$.
Finally, we evaluate $\Delta\chi^2\equiv\sum_j
\Delta\chi^2_j\Theta(\Delta\chi^2_j)$
and accept the light curve for human review provided that
$\Delta\chi^2>\Delta\chi^2_{\rm thresh}$. 

Note that whereas \citet{eventfinder} removed the 10\% worst points,
we are only removing the 5\% worst.  This is because we are already
removing bad-seeing and high-background points as part of our general
procedures.

As for the event finder \citep{eventfinder}, we subtract the 
contributions to $\Delta\chi^2$ of the two
points from each data file that contribute the most to $\Delta\chi^2$
before comparing the result to $\Delta\chi^2_{\rm thresh}$.  This
guards against short-term artifacts.

\subsection{{Group Finding}
\label{sec:group}}

After all candidates have been selected from a patch, a
``friends-of-friends'' algorithm is applied to group together
neighboring light curves that are morphologically similar.
This algorithm is very similar to the one used for the event-finder
\citep{eventfinder} except that the ``friend'' condition is necessarily
different because the ensemble of models is different.  In this case,
(in addition to demanding that the star positions be within 
$4^{\prime\prime}$) we require that 
$|\log [(t_{\rm now}-t_{1,\rm rise})/(t_{\rm now}-t_{2,\rm rise})]|< 0.2$,
where $t_{1,\rm rise}$ and $t_{2,\rm rise}$
are the rise times of the models for the
two candidates.  That is, $|k_1 - k_2|\leq 1$.  The group member with
the highest $\Delta\chi^2$ is the designated as the ``group leader''.
Only the group leader is further considered.

\subsection{{Automated Post-Machine-Selection Vetting}
\label{sec:postselvet}}

The status of many (generally, a substantial majority) of ``group leaders'' 
identified by the grouping algorithm are already known, and hence it
is not necessary to show them to the operator.  That is, they are
already-known microlensing events (which do not require ``rediscovery'')
or they are already-known variables or artifacts.

Candidates that are registered as ``clear microlensing'' 
(see Section~\ref{sec:visual}) during a given year are automatically
suppressed prior to visual review for the rest of the year.  
Similarly, candidates that are registered as ``not microlensing'' 
are also suppressed, in this case for all future years as well.

In addition, all stars that were found to be ``variables'' or ``artifacts''
in any application of the event-finder \citep{eventfinder,2016ef}
from previous years, or to be
a member of a ``group'' whose ``group leader'' was so designated, are
likewise removed from the visual review.

\subsection{{Visual Review}
\label{sec:visual}}

The visual review of events is broadly similar to that of the event-finder
\citep{eventfinder}, but with a number of particular features that are
specific to real-time microlensing alerts.

As with the event finder, the operator is shown a four-panel display
in which the separate light-curve files are aligned via the model parameters,
with the model (a broken line) also shown.
The upper-left panel shows the most recent data, going back twice as
far in time as $t_{\rm rise}$ (but as least 5 days before $t_{\rm rise}$)
and with a range of fluxes that goes $\pm 3\,\sigma$ beyond the range
of the model.  This provides the main information on whether the
data appear consistent with a rising microlensing event.  The
upper-right panel shows the same time range but the vertical axis encompasses
the full range of fluxes observed during this interval.  This is similar
to the event-finder display and allows one to check for 
anomalies, which would be an important signature of microlensing in
cases for which the light curve in the upper-left panel appears
irregular.
The lower-right panel contains the full year of data taken so far with the
same flux range as the upper-left panel.

The biggest difference in this alert display relative to 
the event-finder display
is the lower-left panel.  This shows the earlier-year (currently 2016) data 
aligned according to
the current-year model.  For real microlensing events, this will
usually be flat, and the model-induced alignment will produce
roughly coincident (and flat) light curves for the earlier year.  However, for
artifacts (due for example to bleeds from saturated stars), the alignment 
can be a strong function of the different conditions at
different observatories and so can easily yield misaligned 2016
light curves.  This is an important signature of artifacts.

\begin{figure}
\includegraphics[width=15cm]{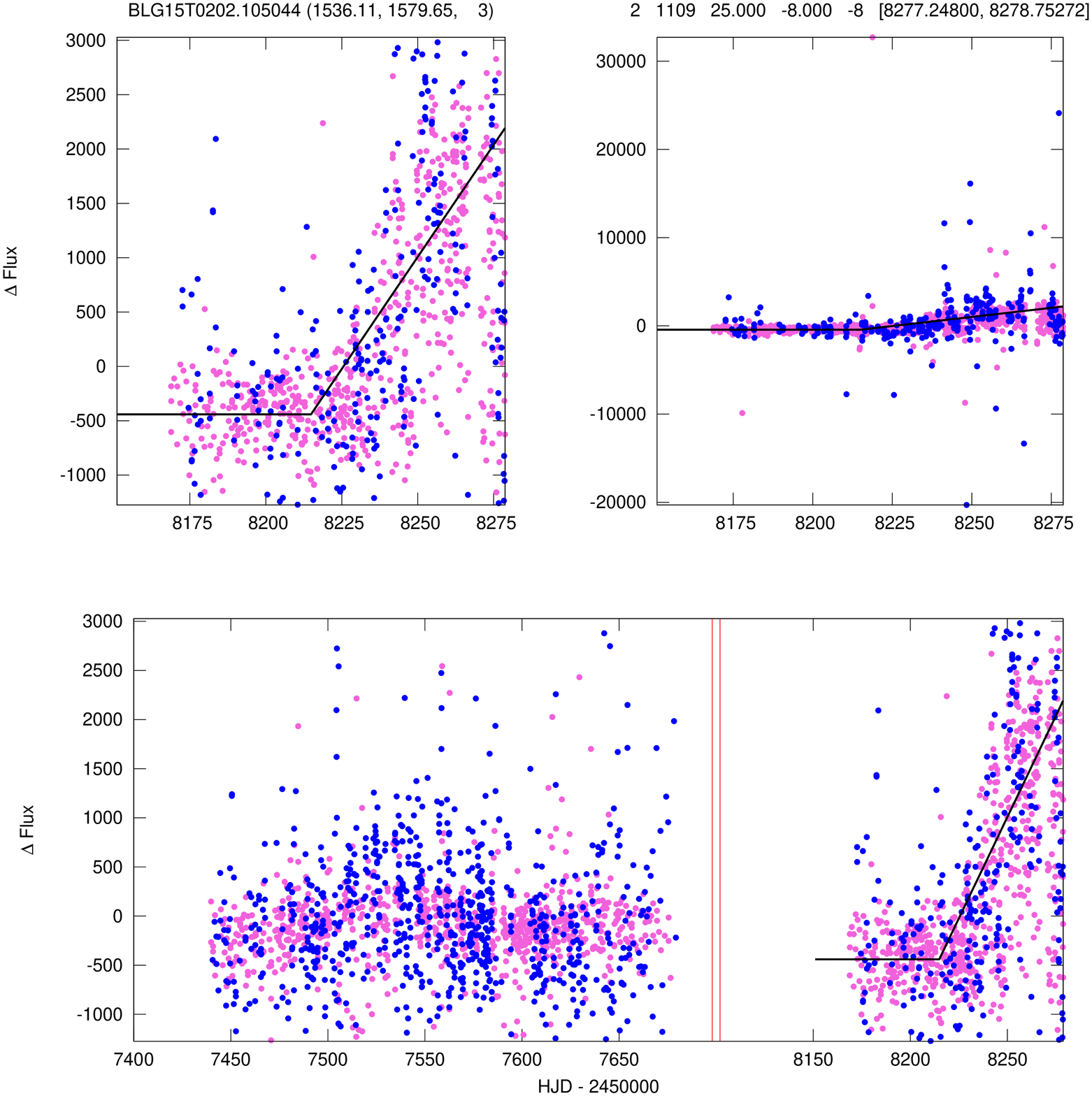}
\caption{Four-panel display of a machine-selected candidate that was rejected by
the operator as an ``artifact''.  The upper-left panel shows a model fit
to the recent data.  The KMTC data contributed the most to $\Delta\chi^2$,
so the candidate name begins ``BLG'' (rather than ``SAO'' or ``SSO'').
The upper-right panel shows the same time frame but is extended vertically
to include all data.  The lower-right panel shows the same vertical range
as the upper-left panel, but for the whole season.  
The lower-left panel shows 2016
data with different observatories aligned according to the 2018 model.
Because KMTC contributes the most to $\Delta\chi^2$, it is shown in
magenta.  KMTS is shown in blue.  KMTA is not shown because it did
not contribute positively to $\Delta\chi^2$.  The operator judged this
to be an ``artifact'' mainly because the high-noise level is typical
of artifacts generated by bleeds from a bright star contaminating
faint stars further down the same column.  The catalog star's 
magnitude $I=25.000$ indicates that it was was too faint in the
underlying DECam catalog \citep{decam} to make a proper estimate
of its true $I$-band magnitude, making it a very plausible ``host''
for such artifacts.  The lower-left panel did not, in this case,
contribute to the ``artifact'' designation.
}
\label{fig:artifact}
\end{figure}

Figure~\ref{fig:artifact} shows the four-panel display for a candidate
that was judged to be an artifact.

All visually-reviewed light curves are given one of the following
four classifications:
\begin{enumerate}
\item[]{C1) Clear microlensing}
\item[]{C2) Probable microlensing}
\item[]{C3) Possible microlensing}
\item[]{C4) Not microlensing}
\end{enumerate}

Events classified as C1 and C2 are added to the web page of 
``public alerts'', with a flag indicating whether they are ``clear''
or ``probable''.
As discussed in Section~\ref{sec:postselvet}, events classified as
C1 and C4, are suppressed for future applications of the program
because the status of these events is considered already known.
On the other hand, future re-appearances of candidates marked
C2 or C3 are labeled as such in the display, as a reminder
in preparation for their possible reclassification based on new
information.  In particular, if a C2 candidate is upgraded
to C1 or downgraded to C3 or C4, then this is marked on the webpage.

\subsection{{Evaluation of Algorithm in Light of Objectives}
\label{sec:eval}}

In Section~\ref{sec:quality}, we outlined three qualitative goals
for the alert system that were derived based on the scientific
objectives and practical constraints discussed in 
Section~\ref{sec:obj+const}.  These were (in permuted order)
``operate as rapidly as possible'', 
``focus on and robustly identify rising events'',
and``minimize the number of false positives'' presented to the operator.

Here, we evaluate the practical implementation described above in
light of these goals.

The most crucial goal is the first: ``operate as rapidly as possible''.
Each of up to $5\times 10^8$ light curves must be reviewed, and these receive
typically 1--100 additional data points per day, with a median of $\sim 8$.
The time-limiting step is the initial review of these $5\times 10^8$ 
light curves,
which the algorithm carries out by checking for at least $N_{\rm high}$
points since the previous review (plus $N_{\rm high}+10$ additional points).
That is, for daily reviews, the processing time is dominated by reading
in 15--100 records, with a median of $\sim 25$.  

This seems modest but it should be compared against another scheme
that is in fact roughly 2--3 times as fast for the case of only one date file
($N_{\rm files}=1$).  In this alternate scheme, each star is assigned
a counter $K_{\rm high}$, which is the number of consecutive high points
ending on the most recent observation.  Then, when a new (for example,
24-hr) set of data are acquired, each successive data point is examined.
If the data point is high, then $K_{\rm high}\rightarrow K_{\rm high}+1$,
and if not, then $K_{\rm high}\rightarrow 0$.  Additional investigation
is then triggered if $K_{\rm high}\geq N_{\rm high}$.  The main time saving
comes from not reading any additional past records.

However, this approach is not easily generalized to the case of
multiple files, in particular coming from multiple observatories.
Ultimately (Section~\ref{sec:ngoodconsec}), we trigger on $N_{\rm high}$
points in any one of $N_{\rm comb}=(2^{N_{\rm files}}-1)$ combinations of files,
i.e., a minimum of seven combinations for $N_{\rm files}=3$.  The
reasons for this will be discussed below, but here we focus on the
implications for speed in the alternate scheme that we are currently discussing.
First, it means that one would require $N_{\rm comb}$ counting parameters
$K_{l,\rm good}, l=1, \ldots N_{\rm comb}$.  Second, it would require time-sorting
the data from all files in order to update these $N_{\rm comb}$ counting
parameters.  Even at the minimum of $N_{\rm comb}=7$, this would be cumbersome.
Moreover, it would become
completely unmanageable when the process is upgraded to combine
overlapping fields, which reaches up to $N_{\rm fields}=4$
($N_{\rm comb} = 4095$) for a considerable number of stars \citep{eventfinder}.
Therefore, such a system would be at most marginally faster under
our initial implementation and would not be upgradable.

The total time for Steps 2 and 3 of the alert-finder algorithm 
is small compared to Step 1
primarily because the fraction of stars that pass Step 1 is small.
In principle, the modeling (Step 3) could take a substantial amount of
time if many models were considered because (in contrast to the
much more numerous event-finder models of \citealt{eventfinder}) the
entire light curve is modeled.  However, because we consider only
16 models, most of the time spent on this step is just reading in
the full light curve files.  This requires, on average, about 30 times
more reads than Step 1.  However, because $< 1\%$ of all light curves 
reach Step 3, the total time spent on this step is still substantially
smaller than on Step 1.

\begin{figure}
\plotone{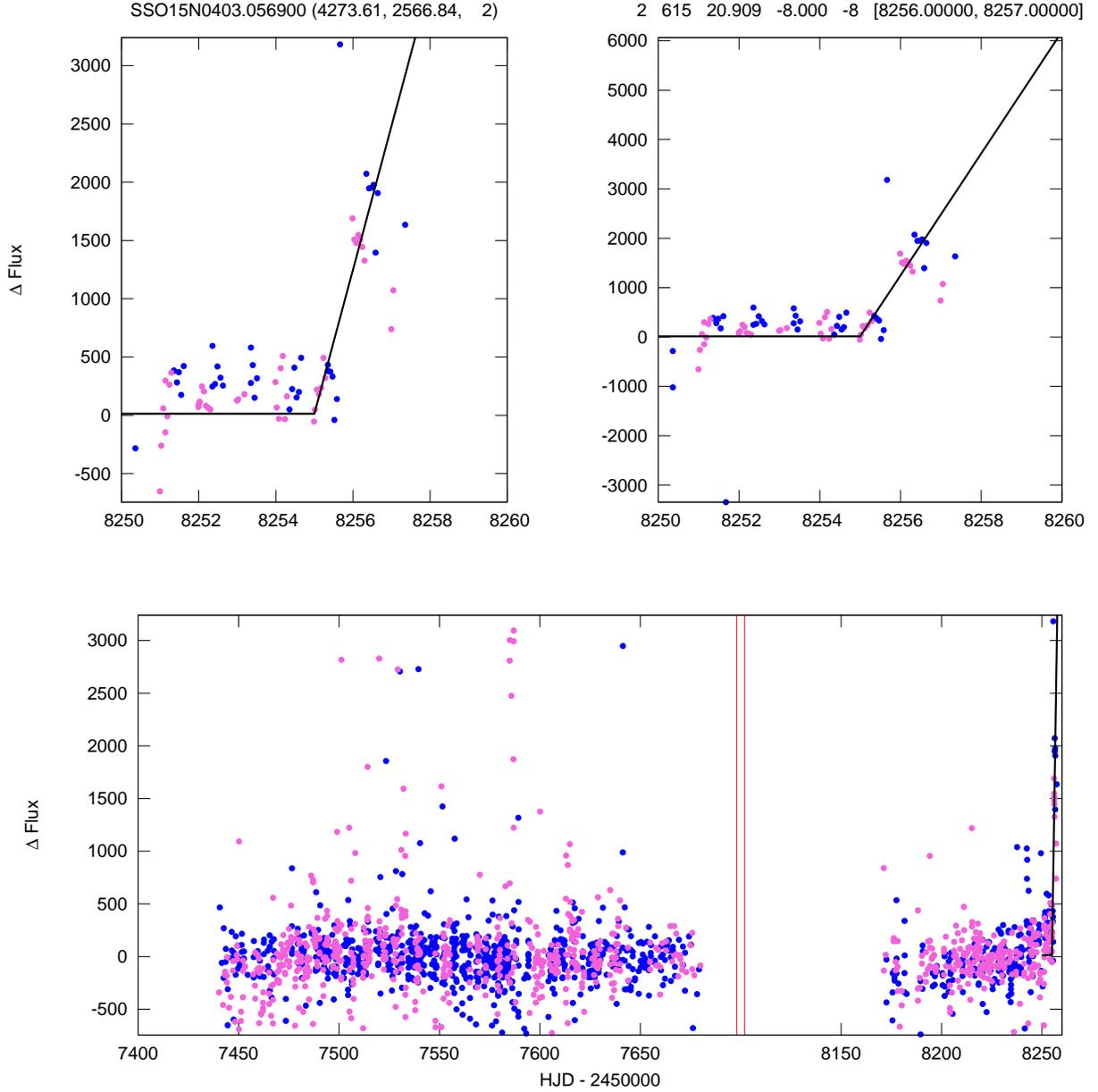}
\caption{Four-panel display of machine-selected candidate that was accepted
by the operator as ``KMT-2018-BLG-0061''.  The panel display is similar
to Figure~\ref{fig:artifact}.
The KMTS data contributed the most to $\Delta\chi^2$,
so the candidate name begins ``SSO'' (rather than ``BLG'' or ``SAO'').
Because the alert-finder algorithm was first run on field BLG15 on 
5 June 2018 (HJD 2458274), the actual selection took place 17 days
after what is shown here.  This can be seen on the alert website.
This figure shows the result of running the alert-finder many times,
each time incrementing $t_{\rm now}$ by one day and setting 
$t_{\rm last}=t_{\rm now}-1$, in order to simulate its regular operation.
This binary event was then first selected on HJD 2458257.  In fact,
at this point, although the light curve 
looks like ``probable'' microlensing, it is
far from obvious that this is a binary microlens.  When it was actually selected
17 days later, this binarity was clear.  The fit on that later date
did not at all
match the binary's features, but $\Delta\chi^2$ was roughly two times
higher than shown here.  See KMT alert website.
}
\label{fig:binary4panel}
\end{figure}

\begin{figure}
\plotone{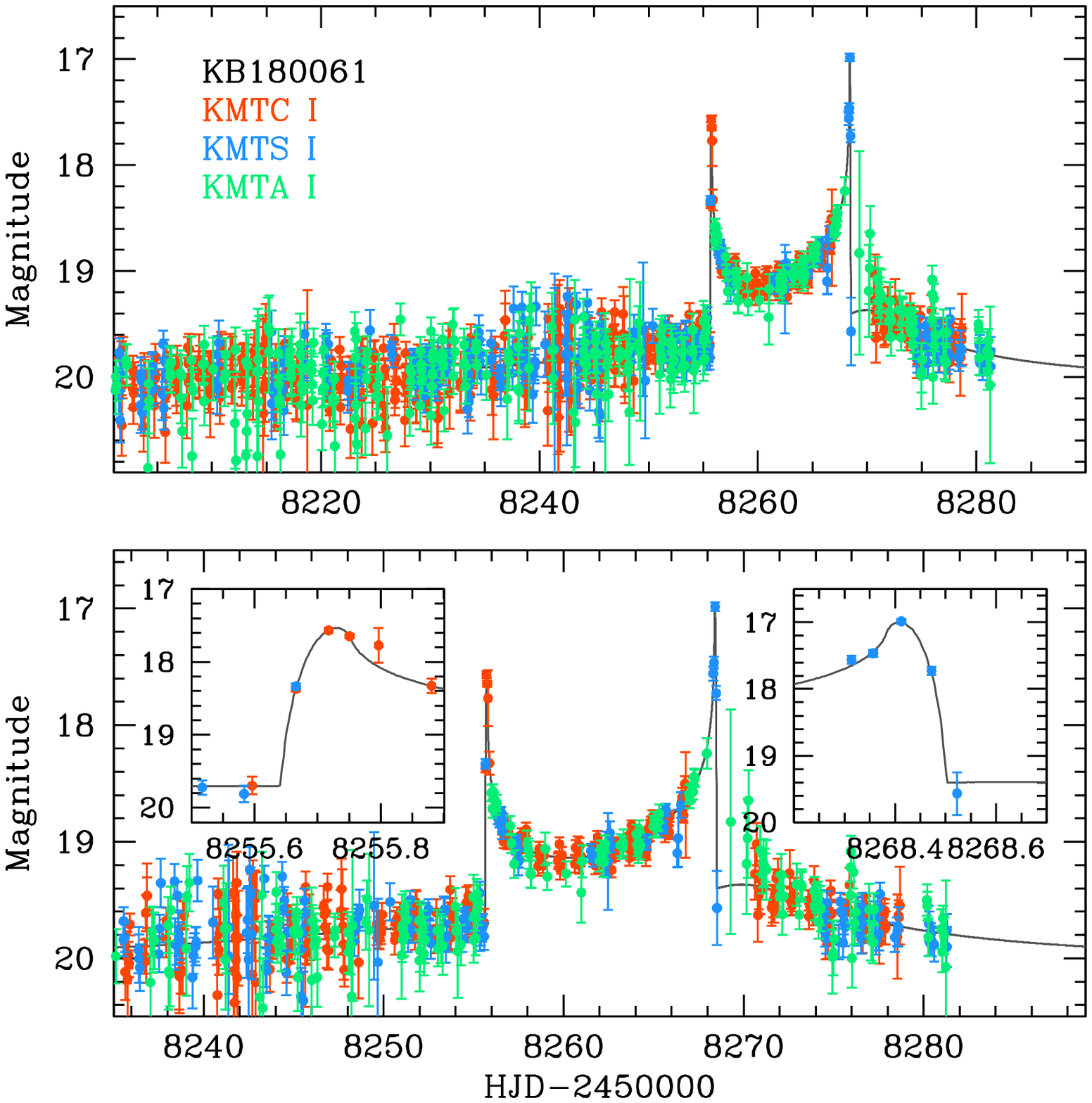}
\caption{Subsequent development of KMT-2018-BLG-0061, whose four-panel
display is shown in Figure~\ref{fig:binary4panel}.  Tender loving care
(TLC) pySIS reductions are shown, which are substantially better
than the automated pySIS-pipeline reductions shown on the KMT alert website.
By comparing this Figure with Figure~\ref{fig:binary4panel}, one can
see that the alert was triggered by data in the trough, and that
the much more dramatic KMTC data on the caustic entrance were actually
``suppressed'' by the algorithm because they were ``outweighed''
by the KMTS and KMTA data in the trough.  Nevertheless, the event
had $\Delta\chi^2=615$, easily exceeding the threshold of
$\Delta\chi^2_{\rm thresh}=400$.
}
\label{fig:binarypysis}
\end{figure}

The next qualitative goal is to 
``focus on and robustly identify rising events''.  
Because the \citet{einstein36} magnification $A(u)=[1 - (u^2/2+1)^2]^{-1/2}$
rises rapidly inside the Einstein radius $(u<1)$ and
quickly goes to $A\rightarrow 1$ outside $(u>1)$, Equation~(\ref{eqn:rise})
is essentially always a reasonable model for rising single-lens 
microlensing events.  Note that it is not necessary for the model to
be a ``good fit'' by some particular standard.  All that is necessary
is that it be substantially better than a flat line.  However, this
model is also quite reasonable for the beginning rise of almost all
binary lenses.  Most of these look qualitatively like single-lens events.
However, even those that become visible by an abrupt jump into a caustic
trough are still much better represented by Equation~(\ref{eqn:rise}) than
by a flat line.  Moreover, in this modeling, the whole light curve is fitted
(with outlier rejection), so that points that grossly deviate from the
pattern (whether due to noise, systematics, or binary-lens signatures)
cannot by themselves cause automated rejection of the candidate.
Figure~\ref{fig:binary4panel} shows the four-panel display corresponding
to the initial selection of binary lens KMT-2018-BLG-0061, while
Figure~\ref{fig:binarypysis} shows the full re-reduced pySIS light-curve
for this event as it subsequently developed.
Hence, Step 3, in itself, is quite robust at retrieving real microlensing
events for visual review.

The key question is then whether real microlensing events will be
rejected at Step 2 ($N_{\rm high}$ consecutive high points).  For events
that rise very quickly, so that suddenly the light curve is much more
than three ``typical sigma'' above the median, essentially all data points
will be ``high'' and it is important to combine all data files in order
to register this as rapidly as possible.  This is particularly true
for short events.  On the other hand, for slowly-rising light curves,
data problems at one or more observatories can easily break consecutive
streaks of high points.  Therefore, the most robust way to deal with
all contingencies is to search for $N_{\rm high}$ consecutive high points
in {\it any} combination of data files.  By practical experimentation,
we found $N_{\rm high}=5$ provides adequate guard against spurious
high points triggering false candidates for $\Gamma\leq 0.4\,{\rm hr}^{-1}$,
but this limit is not adequate for higher-cadence fields.  These
require $N_{\rm high}=10$.  These
restrictions imply that roughly one day of data that is ``significantly
more than $3\,\sigma$ high'' is usually required to trigger Step 3.
And this in turn implies diminishing returns for running the alert system
more than once per day.

\section{{KMTNet Alert Webpage}
\label{sec:web}}

As soon as the visual review of alert candidates is completed on a given
day, the full list of candidate classification is sent to a script
that controls two processes: webpage update and initiating pySIS photometry
\citep{albrow09}.
First, all previously identified candidates are checked for possible
updates in their classification.  Second, new pages are created for each
of the events that are newly classified as ``clear'' or ``probable''
microlensing, and a link to that page is created on the cover page,
which contains a list of the event names, coordinates, classifications,
estimated $I$-band extinction,
and an estimate of the event parameters.  The page initially contains
the 4-panel DIA-based display that led to selection, links to the
underlying DIA data, a finding chart, the event parameters and coordinates,
and cross-identifications with known OGLE and/or MOA events.

Simultaneously, the script places the event on the queue of targets
to obtain pySIS photometry from an automated pipeline.
In contrast to the DIA photometry, which currently is drawn from a single
field, the pySIS photometry includes data from all overlapping KMTNet fields
that contain the event.  As soon as this photometry is complete, 
the page is updated to include a figure showing the pySIS light curve
together with a microlensing model, as well as links to the pySIS
data.  As additional data are taken, the pySIS figure and data files
are updated as soon as they are reduced by the same automated pipeline.

\section{{Ramp-Up of KMTNet Alerts}
\label{sec:ramp}}

KMTNet alerts will be introduced gradually in 2018 according to a
scheme that is fundamentally governed by the availability of 
reduced real-time data.  There are three reasons for adopting
this ramp-up approach, rather than waiting for the system to be
``fully functioning''.  First, alerts for {\it Spitzer} are needed
immediately.  Second, we hope to improve the system by getting feedback
before it is fully functional.  Third, as we discuss below,
it is quite possible that the system will still be only ``mostly functional''
in 2019.  Hence, waiting for ``full function'' could be counterproductive.

As discussed in Section~\ref{sec:const}, 2018 data reduction did not
begin until about 1 May 2018.  Prior to this time, the alert software
was tested on blinded data from 2016 and 2017.  After 1 May, all reduction
efforts were concentrated on five northern bulge fields, BLG14, BLG15, BLG18,
BLG19, and BLG16.  The first four of these are the most
likely to be productive for {\it Spitzer} alerts due to high event rate,
high extinction, and relatively low cadence of OGLE observations.
Those four all have cadence $\Gamma=1\,{\rm hr}^{-1}$, so BLG16 was added
to gain experience with a $\Gamma=0.4\,{\rm hr}^{-1}$ field.  As these
reductions proceeded, more realistic testing and trial runs were done on the
2018 BLG14 data, gradually expanding to a few other fields.
We populated the webpage with all of these ``alerts'',
although no one outside of KMTNet was ``alerted'' at this time.
Following a two weeks of this realistic testing, we made the
page public at 
http://kmtnet.kasi.re.kr/ulens/event/2018/ ,
which includes the above ``alerts''  from the
test phase, as well as all subsequent real alerts.

\subsection{Limitations Due to Data Processing Rate}

For 2018, we will continue to update photometry on all fields that
have previously been ``caught up'', and we will use the 
computer time that remains to ``catch up'' on additional fields,
one at a time.  After each such field is caught up, we will begin
real time alerts on that field.  As we discuss below,
however, we think that it is likely that alert coverage in 2018 will
be mostly restricted to the Northern bulge.

At present the image analysis proceeds at about 57\% of the peak
(i.e., June/July) time-averaged $I$-band data-acquisition rate.  This
(together with the fact that 2018 DIA reductions only began in May) is
the fundamental reason that alerts will be restricted to a modest
fraction of the total KMTNet area in 2018.

As mentioned in Section~\ref{sec:const}, it takes 5 minutes to
do photometry on one image.  Ninety-six processors are assigned to this
process (actually, 24 processors per server for each of four
servers), while 64 other processors
are simultaneously working on the previous step (alignment and subtraction)
for the next set of images.  However, I/O and computational
constraints prevent any wider distribution of the processing.  This
means that the average rate of image processing is one per five minutes,
or 288 per day.  During June and July, we take about 24 $I$-band images
per hour for about 9.4 hours per night at each of the three observatories.  This
is about 675 images per day provided that there is good weather at all sites,
or about 500 images per day allowing for weather and equipment
problems. 

Moreover, in the absence of additional computers
or greater algorithmic efficiency, we will not be able to keep up
with the full data flow at peak season in future years.  
This problem can be significantly ameliorated,
though not completely solved, at relatively low-cost by gradually
curtailing reductions of the three overlap fields BLG41, BLG42, 
and BLG43 beginning about 1 April until these reductions are completely
halted about 1 May.
These three fields consume about 27\% of observation time, but they
only cover about $1\,{\rm deg}^2$ that is not covered by other
fields (primarily BLG01, BLG02, and BLG03).  For the remaining
$11\,{\rm deg}^2$ covered by these fields, the data are very important
for finding and characterizing planets, but are not crucial for
issuing alerts.

This still leaves 3.5 months (1 May to 15 August) when 
real-time data reduction cannot keep pace with data acquisition.
In the absence of other solutions, the approach
would be to leave KMTA data partially (beginning 1 May) or totally
(beginning 1 June) unreduced for this period and to
search for alerts based only on KMTC and KMTS data.

Based on these qualitative considerations, we adopt the following
algorithmic approach to deciding the next image to process from
a possibly accumulating backlog:
\begin{enumerate}
\item[]{D1) Observatory order: KMTC, KMTS, KMTA}
\item[]{D2) Field Order: BLG(14,18,15,19,11,38,16,20,12,02,03,01,04,17,22,35,31,37,21,32,33,34,13,36)}
\item[]{D3) Earliest epoch that is not yet processed}
\item[]{D4) If no unprocessed images from fields BLG01--38, then}
\item[]{\quad D4.1) Observatory order: KMTC, KMTS, KMTA}
\item[]{\quad D4.2) Field Order: BLG(42,43,41)}
\item[]{\quad D4.3) Earliest epoch that is not yet processed}
\end{enumerate}
The field order shown in (D2) prioritizes Northern Bulge
fields, which will continue to be important for {\it Spitzer} alerts
in 2019.  The order could be changed to meet future changes in
priorities.

This algorithm (D1 -- D4)
automatically implements the qualitative features 
described above.  The key question is how it functions in 
practice, in particular during the roughly six weeks when each
observatory can observe about 9.4 hours per night.

The first point is that whenever KMTC is observing (which is a substantial
majority of the time), there will be continuous new data in all
fields, which will enable the selection of new alerts.  That is, because
of condition (D1), KMTC data will always be reduced within a few hours
of the time they are taken.  Second,
if KMTC and KMTS experience intermittent bad weather, then both
observatories will remain approximately caught up.  Then, if KMTC
weather turns bad but KMTS continues to observe, the most recent
KMTS images will be processed within seven hours after they are taken.

The first major ``problem'' with this algorithm occurs if
KMTC and KMTS have extended simultaneous good weather.  For
example, after one week of joint good weather, KMTS would be
four days behind in its processing.  However, due to condition (D2),
this backlog would not be uniform over all fields.  Rather, fields
BLG14, BlG18, BLG15, BLG19, BLG11, BLG38, BLG16, and BLG20 
would be completely caught up,
while the remaining fields would be a week behind.  Hence, new
alerts could be immediately identified from these fields, but not
the others.  Then, as bad weather continued at KMTC, the remaining KMTS
fields would successively catch up in the order specified by (D2).
In the extreme case of all bad weather at KMTC and all good weather
at KMTS, it would take about nine days for KMTS to fully catch up.
However, in more typical scenarios, this catch up would be much
faster.  For example, with two days of good weather at KMTS
followed by two days of bad weather, KMTS would be completely caught up.
Thus, for the great majority of the time that either KMTC or KMTS
is operating, there can be near real-time alerts over all, or 
essentially all fields.

The real difficulty comes when both KMTC and KMTS have simultaneous
bad weather.  Let us consider that this occurs four weeks into
the six-week period of 9.4-hour observability.  Then all KMTA fields
will be at least four weeks behind (and the lower priority fields
more than this).  This dual hiatus would immediately trigger
reductions of KMTA BLG14, which will require about a half day to 
catch up.  Hence, alerts can begin on this field with a half-day delay.
For BLG18, the delay will be one day.  Hence, there would be
a gradual return of alerts beginning with the highest priority
fields.

Thus, at the present time, this algorithm appears to be the
best way to cope with overtaxed computing resources at the
peak of the season.

\subsection{Potential Upgrades}

Finally, we remark on two possible upgrades that are lower priority.

First, we note that two-thirds of the missing $\sim 1\,{\rm deg}^2$
from fields BLG41, BLG42, and BLG43
mentioned above could be recovered by reducing the southern-most
patches $[(1,i), i=1\ldots 8]$ of the two northern chips (K and M)
of fields BLG41, BLG42, and BLG43.  These patches constitute just
1/16 of these fields and so require only about 2\% of the total 
computing time.  However, it is not clear at the present time
whether the modest additional science output warrants the 
added complexity of reducing partial chips.

Second, as we mentioned in Section~\ref{sec:quality}, by focusing
on rising events, we are ignoring not only events that have already
peaked during the year (and so will already have been alerted provided
that they have significant signal) but also events that peaked
between seasons and so may have strong ``purely falling'' signals.
These could, in principle, be recovered by making one or two runs
of a slightly modified version of the same program early in the season.
There would be four modifications, all small. First, one would
replace Equation~(\ref{eqn:rise}) by
\begin{equation}
F(t) = a_0 + a_1 (t-t_{\rm flat})\Theta(-(t-t_{\rm flat})).
\label{eqn:flat}
\end{equation}
Second, one would accept only fits with $a_1<0$.  Third, one would test
a grid of models in which $(t_{\rm flat}-t_{\rm now})$ was both positive
and negative.  Fourth, one would set the ``previous date'' to the beginning
of the season.
It would be quite cumbersome to include such fits
during the 2018 ramp up, but it might be feasible near the beginning
of 2019.


\section{{Disclaimer}
\label{sec:disclaimer}}

As discussed in Section~\ref{sec:obj}, the only scientific objective of
the KMTNet public alerts is to encourage and enable followup
observations that lead to better characterization of microlensing
events discovered by KMTNet, including parallax measurements,
discovery of new planets, and improved characterization of planets
and binaries that are already recognizable.

We welcome cooperation with others who obtain such followup data,
as well as with other surveys that obtain data on the same events.
However, the KMTNet data that are displayed and linked on these pages
are completely proprietary and cannot be published or reproduced 
in talks, conference proceedings, etc, without the express permission
of KMTNet.  As discussed by \citet{eventfinder,2016k2,2016ef},
all KMTNet light curves
will eventually be made public in a series of annual data releases.
Our goal is that these releases should be made within well under a year
of the end of each season.  However, until the date of release, as specified
in the data-release papers, these data remain completely proprietary.

\acknowledgments 
Work by AG was supported by AST-1516842 from the US NSF.
IGS and AG were supported by JPL grant 1500811.
“Work by C. Han was supported by the grant
(2017R1A4A1015178) of National Research Foundation
of Korea.
This research has made use of the KMTNet system operated by the Korea
Astronomy and Space Science Institute (KASI) and the data were obtained at
three host sites of CTIO in Chile, SAAO in South Africa, and SSO in
Australia.

\end{document}

%% file: author.tex
 \author{\textsc{
Hyoun-Woo Kim$^{1}$, 
Kyu-Ha Hwang$^{1}$, 
Yossi Shvartzvald$^{2}$, 
Jennifer C. Yee$^{3}$, 
Michael D.Albrow$^{4}$, 
Sang-Mok Cha$^{1,5}$, 
Sun-Ju Chung$^{1,6}$, 
Andrew Gould$^{1,7,8}$, 
Cheongho Han$^{9}$, 
Youn Kil Jung$^{1}$, 
Dong-Jin Kim$^{1}$,
Seung-Lee Kim$^{1,6}$, 
Chung-Uk Lee$^{1,6}$, 
Dong-Joo Lee$^{1}$,
Yongseok Lee$^{1,5}$, 
Byeong-Gon Park$^{1,6}$, 
Richard W. Pogge$^{8}$
Yoon-Hyun Ryu$^{1}$, 
In-Gu Shin$^{3}$, 
Weicheng Zang$^{10}$, } }

\affil{$^{1}$Korea Astronomy and Space Science Institute, Daejon
34055, Republic of Korea}

\affil{$^{2}$IPAC, Mail Code 100-22, Caltech, 1200 E. California Blvd., 
Pasadena, CA 91125, USA}

\affil{$^{3}$ Harvard-Smithsonian Center for Astrophysics, 60 Garden
St., Cambridge, MA 02138, USA}

\affil{$^{4}$University of Canterbury, Department of Physics and
Astronomy, Private Bag 4800, Christchurch 8020, New Zealand}

\affil{$^{5}$School of Space Research, Kyung Hee University,
Yongin, Kyeonggi 17104, Republic of Korea}


\affil{$^{6}$Korea University of Science and Technology, 
217 Gajeong-ro, Yuseong-gu, Daejeon 34113, Republic of Korea}

\affil{$^{7}$Max-Planck-Institute for Astronomy, K\"{o}nigstuhl 17,
69117 Heidelberg, Germany}

\affil{$^{8}$Department of Astronomy, Ohio State University, 140 W.
18th Ave., Columbus, OH 43210, USA}

\affil{$^{9}$Department of Physics, Chungbuk National University,
Cheongju 28644, Republic of Korea}

\affil{$^{10}$Physics Department and Tsinghua Centre for
Astrophysics, Tsinghua University, Beijing 100084, China}



